\documentclass[11pt]{article}
\usepackage{epsf}
\begin{document}
\begin{flushright}
LPT-ORSAY-0047 
\end{flushright}
\vskip .6cm
\begin{center}
{\large {\bf Cosmological constant vs. quintessence}}\footnote{ Lectures
given at Les Houches summer school ``The early Universe'', July 1999 and
Peyresq 4 meeting.}
\vskip .5cm
%\today
\vskip 1cm
Pierre Bin\'etruy, LPT, Universit\'e Paris-Sud, France
\vskip 2cm
{\bf Abstract}
\end{center}
\vskip 1cm
There is some  evidence that the  Universe is presently undergoing 
accelerating expansion. This has restored some credit to the scenarios with
a non-vanishing cosmological constant. From the point of view of a
theory of fundamental interactions, one may argue that a dynamical
component with negative pressure is easier to achieve. As an
illustration, the quintessence scenario is described and its
shortcomings are discussed in connection with the nagging
``cosmological constant problem''. 
\vskip 2cm

\section{Cosmological constant}

 As is well known, the cosmological constant appears as a 
constant  in
the Einstein equations~: 
\begin{equation} R_{\mu \nu} - {1 \over 2}
g_{\mu \nu} R = - 8 \pi G_N T_{\mu \nu} + \lambda g_{\mu \nu},
\label{Einstein} 
\end{equation} 
where $G_N$ is Newton's constant and
$T_{\mu \nu}$ is the energy-momentum tensor. The cosmological constant
$\lambda$ is thus of the dimension of an inverse length squared.  It was
introduced by Einstein \cite{Einstein,rev} in order to build a static 
universe model, its
repulsive effect compensating the gravitational attraction, but, as we
now see, constraints on the expansion of the Universe impose for it
a very small upper value.

It is more convenient to work in the  specific context of a
Friedmann universe, with a Robertson metric~:
\begin{equation}
ds^2 = dt^2 - a^2(t) \left[ {dr^2 \over 1-kr^2} + r^2 \left( 
d\theta^2 + \sin^2 \theta d\phi^2 \right) \right], \label{RW}
\end{equation}
where $a(t)$ is the cosmic scale factor. Implementing energy
conservation into the Einstein equations then leads to the Friedmann
equation which gives an expression for the Hubble parameter $H$~:
\begin{equation}
H^2 \equiv {{\dot a}^2 (t)  \over a^2 (t)} = {1\over 3} \left( \lambda
+ 8 \pi G_N \rho \right) - {k \over a^2}, \label{Friedmann}
\end{equation}
where, using standard notations, $\dot a$ is the time derivative of
the cosmic scale factor, $\rho= {T^0}_0$ is the energy density
and the term proportional to $k$ is a spatial curvature term
(see (\ref{RW})). Note that the cosmological constant appears  as a
constant contribution to the Hubble parameter.

Evaluating each term of the Friedmann equation at present time allows
for a rapid estimation of an upper limit  on $\lambda$. Indeed, we
have $H_0 = h_0 \times 100 \; {\rm km.s}^{-1}{\rm Mpc}^{-1}$ with
$h_0$ of order one, whereas  the present energy density $\rho_0$ is
certainly within one order of magnitude of the critical energy density
$\rho_c = 3 H_0^2 / (8 \pi G_N)= h_0^2 \ 2.10^{-26}$ kg.m$^{-3}$;
moreover the spatial curvature term certainly does not represent
presently a dominant  contribution to the expansion of the Universe.
Thus, (\ref{Friedmann}) implies    the following constraint on
$\lambda$~: \begin{equation}
|\lambda| \le H_0^2.
\end{equation}
In other words, the length scale $\ell_{\Lambda} \equiv
|\lambda|^{-1/2}$ associated with the cosmological constant must be
larger than $H_0^{-1}=h_0^{-1}. 10^{26}$ m, and thus a macroscopic
distance. 

This is not a problem as long as one remains classical. Indeed,
$H_0^{-1}$ provides a natural macroscopic scale for our present
Universe. The problem arises when one tries to combine gravity with
the quantum theory. Indeed, from the Newton's constant {\em and} the
Planck constant $\hbar$ one can construct a mass scale or a length
scale \begin{eqnarray}
m_P &=& \sqrt{{\hbar c \over 8 \pi G_N}}=2.4 \times 10^{18} \; {\rm
GeV/c}^2, \nonumber \\
\ell_P &=& {\hbar \over m_P c} = 8.1 \times 10^{-35} \; {\rm m}
\nonumber
\end{eqnarray}
The above constraint now reads~:
\begin{equation}
\ell_{\Lambda} \equiv |\lambda|^{-1/2} \ge {1 \over H_0} \sim
10^{60} \ \ell_P. \label{lambdaconstr}
\end{equation}
In other words, there are more than sixty orders of magnitude
between the scale associated with the cosmological constant and
the scale of quantum gravity.   
 
A rather obvious solution is to take $\lambda = 0$. This is as valid
a choice as any other in a pure gravity theory. Unfortunately, it is
an unnatural one when one introduces any kind of matter. Indeed,
set $\lambda$ to zero but assume that there is a non-vanishing vacuum
({\em i.e.} groundstate) energy: $<T_{\mu\nu}> = - <\rho>
g_{\mu\nu}$, then the Einstein equations (\ref{Einstein}) read
\begin{equation}
R_{\mu \nu} - {1 \over 2} g_{\mu \nu} R = - 8 \pi G_N T_{\mu \nu}
+ 8 \pi G_N <\rho > g_{\mu \nu}, \label{vacenergy}
\end{equation}
The last term is interpreted as an effective cosmological constant~:
\begin{equation}
\lambda_{eff} = 8 \pi G_N <\rho > \equiv {\Lambda^4 \over m_P^2}.
\label{lambdaeff}
\end{equation}
Generically, $<\rho >$ receives a non-zero contribution from symmetry
breaking: for instance, the scale $\Lambda$ would be typically of the
order of $100$ GeV in the case of the gauge symmetry breaking of the 
Standard Model or  $1$ TeV in the case of supersymmetry breaking.
But the constraint (\ref{lambdaconstr}) now reads:
\begin{equation}
\Lambda \le 10^{-30} \; m_P \sim 10^{-3} \ {\rm eV}.
\label{Lambdaconstr}
\end{equation}
It is this very unnatural fine-tuning of parameters (in explicit
cases $<\rho >$ and thus $\Lambda$ are functions of the parameters of
the theory) that is referred to as the {\em cosmological constant problem}, or
more accurately the vacuum energy problem.

\section{The role of supersymmetry}

If the vacuum energy is to be small, it may be easier to have it
vanishing through some symmetry argument. Global supersymmetry is
the obvious candidate. Indeed, the supersymmetry algebra
\begin{equation}
\{ Q_r , \bar Q_s \} = 2 \gamma^\mu_{rs} P_\mu \label{SUSY}
\end{equation}
yields the following relation between the Hamiltonian $H=P_0$ and the
supersymmetry generators $Q_r$: 
\begin{equation}
H = {1 \over 4} \sum_r Q_r^2,
\end{equation}
and thus the vacuum energy $<0|H|0>$ is vanishing if the vacuum is
supersymmetric ($Q_r |0> \ = 0$). 

Unfortunately, supersymmetry has to be broken at some scale since
its prediction of equal mass for bosons and fermions is not observed
in the physical spectrum. Then $\Lambda$ is of the order of the
supersymmetry breaking scale, that is a few hundred GeV to a TeV. 

However, the right framework to discuss these issues is supergravity
{\em i.e.}  local supersymmetry since  locality implies here,
through the algebra (\ref{SUSY}), invariance under ``local''
translations that are the diffeomorphisms of general relativity.
In this theory, the graviton, described by the linear perturbations of
the metric tensor $g_{\mu \nu}(x)$, is associated through
supersymmetry with a spin 3/2 field, the gravitino $\psi_\mu$.
One may write a supersymmetric invariant combination of terms in the
action:
\begin{equation} 
{\cal S} = \int d^4x \sqrt{g} \left[ 3 m_P^2 m_{3/2}^2 - m_{3/2}
\bar \psi_\mu \sigma^{\mu \nu} \psi_\nu \right],
\end{equation}
where $\sigma_{\mu\nu} = [\gamma_\mu, \gamma_\nu]/4$. If the first
term is made to cancel the vacuum energy, then the second term is
interpreted as a mass term for the gravitino. We thus see that the
criterion for spontaneous symmetry breaking changes from global
supersymmetry ({\em non-vanishing vacuum energy}) to local supersymmetry
or supergravity ({\em non-vanishing gravitino mass}). It is somewhat a
welcome news that a vanishing vacuum energy is not tied to a
supersymmetric spectrum. On the other hand, we have lost the only
rationale that we had to  explain a zero cosmological constant.

Let us recall for future use that, in supergravity, the potential for 
a set of scalar fields $\phi^i$ is written in terms of the K\"ahler 
potential $K(\phi^i, \bar \phi^{\bar i})$ (the normalisation of the scalar 
field kinetic terms is simply given by the K\"ahler metric $K_{i \bar j} 
= \partial^2 K / \partial \phi^i \partial \bar \phi^{\bar j}$) and of the superpotential 
$W(\phi^i)$, an holomorphic function of the fields:
\begin{equation}
V = e^{K/m_P^2} \left[ \left(W_i + {K_i \over m_P^2} W \right) K^{i \bar j} 
\left( \bar W_{\bar j} + {\bar K_{\bar j} \over m_P^2} \bar W \right) 
- 3 {|W|^2 \over m_P^2} \right] + \; {\rm D} \; {\rm terms}
\label{sugrapot}
\end{equation}
where $K_i = \partial K / \partial \phi^i$, etc. and $K^{i \bar j}$ is 
the inverse metric of $K_{i \bar j}$. Obviously, the positive definiteness of 
the global supersymmetry scalar potential is lost in supergravity.

\section{Observational results}

Over the last years, there has been an increasing number of
indications that the Universe is presently undergoing accelerated
expansion. This appears to be a strong departure from the standard
picture of a matter-dominated Universe. Indeed, the standard equation
for the conservation of energy,
\begin{equation}
\dot \rho = - 3 (p+ \rho) H , \label{encons}
\end{equation}
allows to derive from the Friedmann equation (\ref{Friedmann}), written 
in the case of a universe dominated by a component with energy density
$\rho$ and pressure $p$~:
\begin{equation}
{\ddot a \over a} = -{4 \pi G_N \over 3} (\rho + 3 p). \label{accdec}
\end{equation}
Obviously, a matter-dominated ($\rho \sim 0$) universe is decelerating.
One needs instead a component with a negative pressure.

A cosmological constant is associated with a contribution to the
energy-momentum tensor as in (\ref{vacenergy})(\ref{lambdaeff}): 
\begin{equation}
T^\mu_\nu = - \Lambda^4 \delta^\mu_\nu = (-\rho, p,p,p)
\end{equation}
The associated equation of motion is therefore 
\begin{equation}
p= -\rho. \label{eqlambda}
\end{equation}
It follows from (\ref{accdec}) that a cosmological constant  tends to
accelerate expansion. 

The discussion of data is thus often expressed in
terms of the energy density $\Lambda^4$ stored in the vacuum versus the
energy density $\rho_M$ in matter fields (baryons, neutrinos, hidden
matter,...). It is customary to normalize with the
critical density (corresponding to a flat Universe)~:
\begin{equation}
\Omega_{\Lambda}={\Lambda^4 \over \rho_c}, \; \;
\Omega_M={\rho_M \over \rho_c}, \; \; \; \; 
\rho_c={3 H^2 \over 8 \pi G_N}.
\end{equation}
The relation
\begin{equation}
\Omega_M + \Omega_{\Lambda} = 1,
\end{equation}
a prediction of many inflation scenarios, is found to be compatible
with recent Cosmic Microwave Background measurements
\cite{Boomerang}\footnote{This follows from the fact that the first
acoustic peak is expected at an ``angular'' scale $\ell \sim
200/\sqrt{\Omega_M + \Omega_\Lambda}$ \cite{KSS}.}. It is striking that
independent methods based on the measurement of different observables on
rich clusters of galaxies all point towards a low value of $\Omega_M
\sim 1/3$ \cite{BOPS}: mass-to-light ratio, baryon mass to total cluster
mass ratio (the total baryon density in the Universe being fixed by
primordial nucleosynthesis), cluster abundance. This necessarily implies
a non-vanishing $\Omega_{\Lambda}$ (non-vanishing cosmological constant
or a similar dynamical component).

There are indeed some indications going in this direction from
several types of observational data. One which has been much
discussed lately uses supernovae of type Ia as standard
candles\footnote{ by calibrating them according to the timescale of
their brightening and fading.}. Two groups, the Supernova Cosmology
Project \cite{SCP} and the High-$z$ Supernova Search \cite{HZS} have
found that distant supernovae appear to be fainter than expected in a
flat matter-dominated Universe. If this is to have
a cosmological origin, this means that, at fixed redshift, they are at
larger distances than expected in such a context 
and thus that the Universe is accelerating. 

More
precisely, the relation between the flux $f$ received on earth and the
absolute luminosity ${\cal L}$ of the supernova depends on its redshift
$z$, but also on the geometry of spacetime. Traditionally, flux and
absolute luminosity are expressed  on a log scale as apparent magnitude
$m_B$  and absolute magnitude $M$ (magnitude is $-2.5 \ \log_{10}$
luminosity + constant). The relation then reads 
\begin{equation} 
m_B = 5\log (H_0d_L) + M - 5 \log H_0 + 25. 
\end{equation} 
The last terms are $z$-independent,  {\em if one assumes that supernovae
of type Ia are standard candles}; they are then measured by using low $z$
supernovae. The first term, which involves the luminosity distance $d_L$,  
varies logarithmically with $z$ up to
corrections which depend on the geometry. Expanding in $z$ \footnote{ Of
course, since supernovae of redshift $z \sim 1$ are now being observed,
an exact expression \cite{CPT} must be used to analyze data. The more
transparent form of (\ref{dLsimple}) gives the general trend.}, one
obtains \cite{Book}~: 
\begin{equation} 
H_0 d_L = c z \left[ 1 + {1-q_0 \over 2} z + \cdots \right], \label{dLsimple}
\end{equation} 
where $q_0\equiv - a {\ddot a} / {\dot a}^2$ is the
deceleration parameter.  This parameter is easily obtained from
(\ref{accdec}): in a spatially flat Universe with only matter and a
cosmological constant ({\em cf}. (\ref{eqlambda})), $\rho = \rho_M +
\Lambda^4$ and $p = - \Lambda^4$  which gives
\begin{equation}
q_0 = \Omega_M / 2 - \Omega_{\Lambda}. \label{q0}
\end{equation}

This allows to put some limit on $\Omega_{\Lambda}$ on the model
considered here (see Fig. 1). 
\begin{figure}[htb] 
\centerline{\epsfxsize=12cm\epsfbox{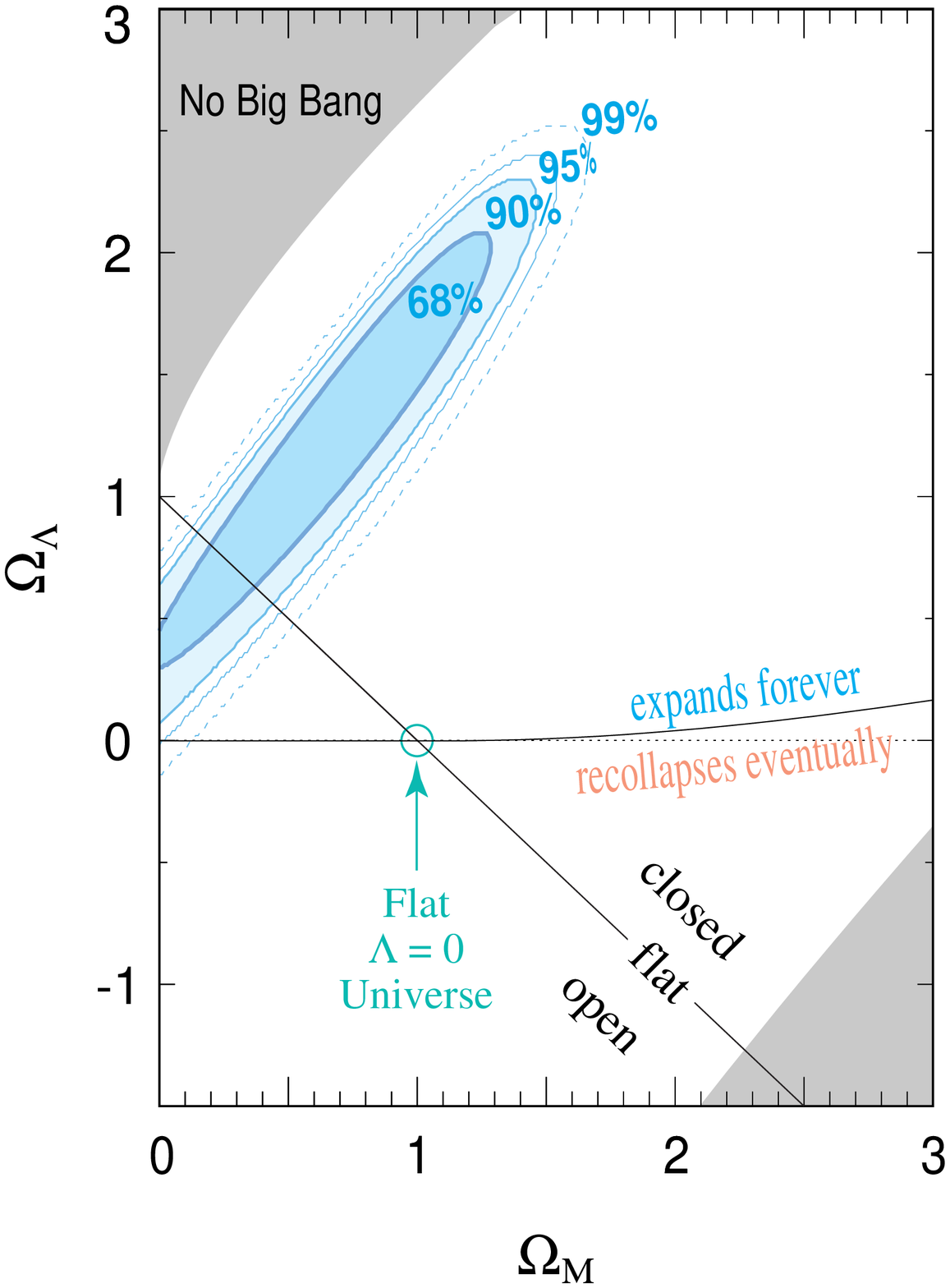}}
\caption{ Best-fit coincidence regions in the $\Omega_M-\Omega_\Lambda$ 
plane, based on the analysis of 42 type Ia supernovae discovered by the 
Supernova Cosmology Project \cite{SCP}. } 
\label{fig1} 
\end{figure}

Let us note that the combination
(\ref{q0}) is `orthogonal' to the combination $\Omega_T \equiv \Omega_M
+ \Omega_\Lambda$ measured in CMB experiments (see footnote preceding
page). The two measurements are therefore complementary: this is
sometimes referred to as `cosmic complementarity'. 

Of course, such type of measurement is sensitive to many possible
systematic effects (evolution besides the light-curve timescale
correction, etc.), and this  has fueled a healthy debate on the 
significance of  present data. This obviously requires  more statistics
and improved quality of spectral measurements. A particular tricky
systematic effect is the possible presence of dust that would dimmer
supernovae at large redshift.

Other results come from gravitational lensing. The deviation of light
rays by an accumulation of matter along the line of sight depends on
the distance to the source \cite{Book}
\begin{equation}
r=\int_t^{t_0} {dt \over a(t)} = {1 \over a(t_0) H_0}
\left( z - {1\over 2} (1+q_0)z^2 + \cdots \right) \label{distance}
\end{equation}
and thus on the cosmological parameters $\Omega_M$ and
$\Omega_\Lambda$. As $q_0$ decreases ({\em i.e.} as the Universe
accelerates), there is more volume and more lenses between the observer
and the object at redshift $z$. Several methods are used: abundance of
multiply-imaged quasar sources \cite{Kochanek}, strong lensing by
massive clusters of galaxies (providing multiple images or arcs)
\cite{sl}, weak lensing \cite{Bernardeau}.

\section{Quintessence} 

From the point of view of high energy physics, it is however
difficult to imagine a rationale for a pure cosmological constant,
especially if it is nonzero but small compared to the typical
fundamental scales (electroweak, strong, grand unified or Planck
scale).  There should be physics
associated with this form of energy and therefore dynamics. 
For example, in the context of string models, any dimensionful
parameter is expressed in terms of the fundamental string scale $M_s$
and vacuum expectation values of scalar fields. The physics of the
cosmological constant is then the physics of the corresponding  scalar
fields.

Introducing
dynamics generally modifies the equation of state (\ref{eqlambda}) to
the more general form with negative pressure~: 
\begin{equation}
p = w \rho, \; \; w < 0. \label{eqw}
\end{equation}
Let us recall that $w=0$ corresponds to non-relativistic matter
(dust) whereas $w=1/3$ corresponds to radiation. A network of light,
nonintercommuting topological defects \cite{Vilenkin,SP} on
the other hand gives $w=-n/3$ where $n$ is the dimension of the
defect {\em i.e.} $1$ for a string and $2$ for a domain wall. Finally, 
the equation of state for a minimally coupled scalar field necessarily 
satisfies the condition 
$w \ge -1$. 

Experimental data may constrain such a dynamical component just as it
did with the cosmological constant. For example, in a spatially flat
Universe with only matter and an unknown component $X$ with equation of
state $p_X = w_X \rho_X$, one obtains from (\ref{accdec}) with
$\rho=\rho_M+\rho_X$, $p=w_X \rho_X$ the following form for the
deceleration parameter
\begin{equation}
q_0 = {\Omega_M \over 2} + (1 + 3 w_X) {\Omega_X \over 2},
\end{equation}
where $\Omega_X=\rho_X/\rho_c$. Supernovae results
give a constraint on the parameter $w_X$ as shown in Fig. 2. 
\begin{figure}[htb] 
\centerline{\epsfxsize=12cm\epsfbox{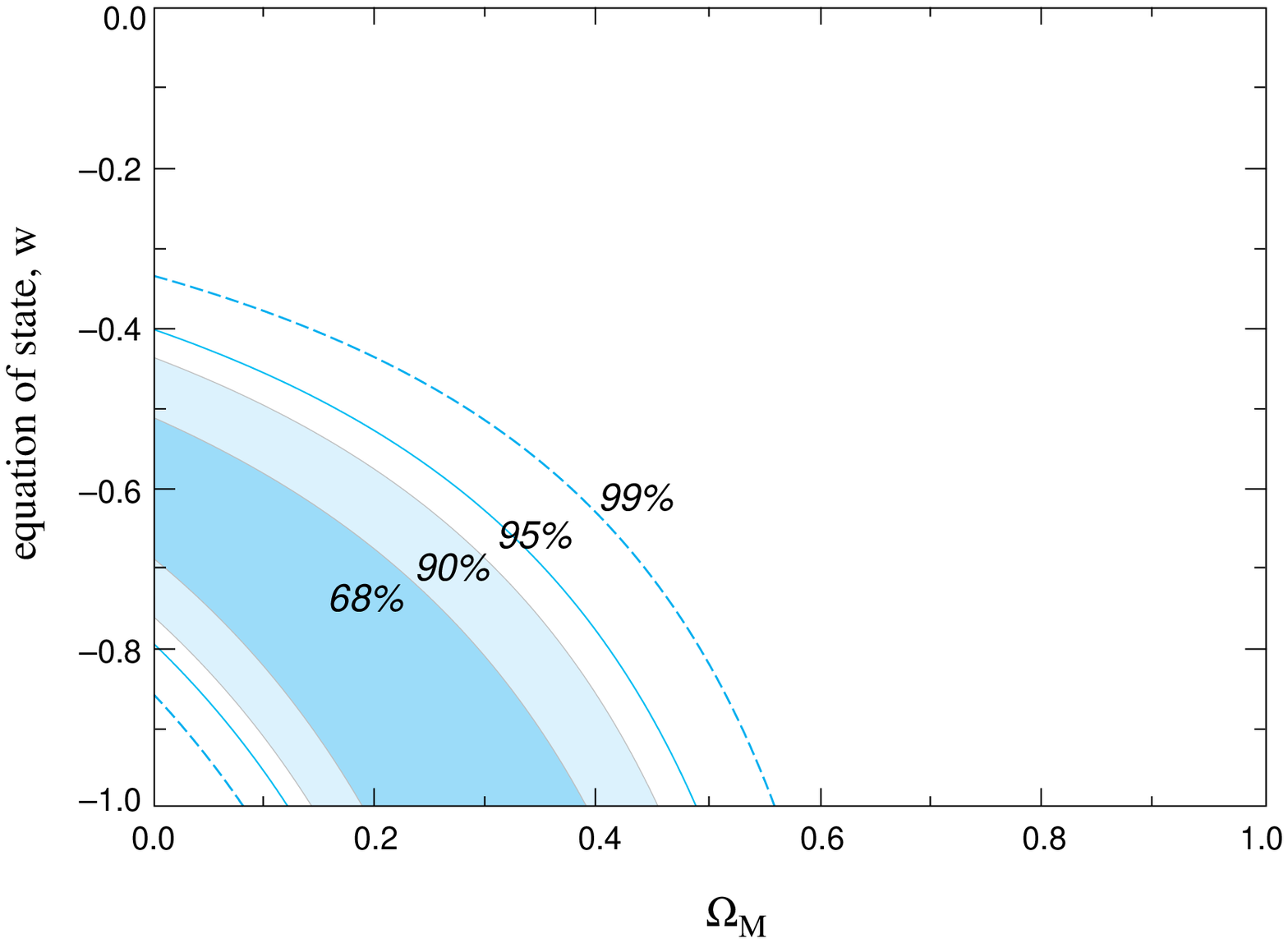}}
\caption{ Best fit coincidence regions in the $\Omega_M-w_X$ plane for
an additional energy density component $\Omega_X$ with equation of state 
$w_X=p_X/\rho_X$, assuming flat cosmology ($\Omega_M+\Omega_X=1$); based on 
the same analysis \cite{SCP} as in Fig.1. } 
\label{fig2} 
\end{figure}
Similarly,
gravitational lensing effects are sensitive to this new component
through (\ref{distance}).

A particularly interesting candidate in the context of fundamental
theories is the case of a scalar\footnote{ A vector field or any field
which is not a Lorentz scalar must have settled down to a vanshing
value. Otherwise, Lorentz invariance would be spontaneously broken.} 
field $\phi$ slowly evolving in a
runaway potential which decreases monotonically to zero as $\phi$ goes to
infinity \cite{Wett,RP,CDS}. This is often referred to as {\em
quintessence}. This can be extended to the case of a
very  light field (pseudo-Goldstone boson) which is presently
relaxing to its vacuum state \cite{FHSW}. We will discuss the two
situations in turn.

\subsection{Runaway quintessence}

A runaway potential is frequently present in
models where supersymmetry is dynamically broken. Indeed,
supersymmetric theories are characterized by a scalar potential with
many flat directions, {\em i.e.} directions $\phi$ in field space for
which the potential vanishes. The corresponding degeneracy is lifted
through dynamical supersymmetry breaking, that is supersymmetry
breaking through strong interaction effects. In some instances (dilaton
or compactification radius), the field expectation value $<~\phi~>$ 
actually provides the value of the strong interaction coupling.
Then at infinite $\phi$ value, the coupling effectively goes to zero
together with the supersymmetry breaking effects and the flat
direction is restored: the potential decreases monotonically to
zero as  $\phi$ goes to infinity.

Dynamical supersymmetry breaking scenarios are often advocated
because they easily yield the large scale hierarchies necessary in
grand unified or superstring theories in order to explain the
smallness of the electroweak scale with respect to the fundamental
scale. Let us take the example of supersymmetry  breaking by gaugino 
condensation in effective superstring theories. The value $g_0$ of the
gauge coupling at the string scale $M_s$ is provided by the vacuum
expectation value of the dilaton field $s$ (taken to be dimensionless by
dividing by $m_P$) present among the massless
string modes: $g_0^2 = <s>^{-1}$. If the gauge group has a one-loop
beta function coefficient $b$, then the running gauge coupling
becomes strong at the scale
\begin{equation}
\Lambda \sim M_s e^{-1/2b g^2_0} = M_s e^{-s/2b}. \label{Lambda}
\end{equation}
At this scale, the gaugino fields are expected to condense. Through 
dimensional analysis, the gaugino condensate $<\bar \lambda \lambda>$
is expected to be of order $\Lambda^3$. Terms quadratic in the
gaugino fields thus yield in the effective theory below condensation
scale a potential for the dilaton~:
\begin{equation}
V  \sim \left| <\bar \lambda \lambda > \right|^2 \propto e^{-3s/b}.
\label{larges}
\end{equation}
The $s$-dependence of the potential is of course more complicated and
one usually looks for stable minima with vanishing cosmological
constant. But the behavior (\ref{larges}) is characteristic of the
large $s$ region and  provides a potential slopping down to zero at
infinity as required in the quintessence solution. A similar behavior
is observed for moduli fields whose {\em vev} describes the radius of
the compact manifolds which appear from the compactification from 10
or 11 dimensions to 4 in superstring theories \cite{B}.

Let us take therefore the example of an exponentially decreasing
potential.  More explicitly, we consider the following
action
\begin{equation}
{\cal S}= \int d^4x \sqrt{g} \left[ -{m_P^2 \over 2} R - { 1 \over 2}
\partial^\mu \phi \partial_\mu \phi - V(\phi) \right], \label{action}
\end{equation}
which describes a real scalar field $\phi$ minimally coupled with
gravity and the self-interactions of which are described by the
potential:
\begin{equation}
V(\phi) = V_0 e^{-\lambda \phi/m_P}, \label{Vexp}
\end{equation}
where $V_0$ is a positive constant.

 The energy density and pressure stored in the scalar field are
respectively~:
\begin{equation}
\rho_\phi = {1 \over 2} \dot \phi^2 + V( \phi), \; \;
p_\phi = {1 \over 2} \dot \phi^2 - V( \phi). \label{rhoandp}
\end{equation}
We will assume that the  background (matter and radiation) energy
density  $\rho_B$  and pressure $p_B$ obey a standard equation of
state 
\begin{equation}
p_B = w_B \rho_B.
\end{equation}

If one neglects the spatial curvature ($k \sim 0$), the equation of
motion for $\phi$ simply reads
\begin{equation}
\ddot \phi + 3 H \dot \phi = - {dV \over d \phi}, \label{eomphi}
\end{equation}
with 
\begin{equation}
H^2 = {1 \over 3 m_P^2} ( \rho_B + \rho_\phi). \label{H2}
\end{equation}

This can be rewritten as
\begin{equation}
\dot \rho_\phi  = -3 H  {\dot \phi}^2. \label{eomrhophi} 
\end{equation}
We are looking for {\em scaling solutions i.e.} solutions where the
$\phi$ energy density scales as a power of the cosmic scale factor: 
$\rho_\phi \propto a^{-n_\phi}$ or $\dot \rho_\phi / \rho_\phi
= - n_\phi H$. In this case, one easily obtains from
(\ref{rhoandp}) and (\ref{eomrhophi}) that the $\phi$ field obeys a
standard equation of state  \begin{equation} 
p_\phi = w_\phi \rho_\phi,
\end{equation}
with 
\begin{equation}
w_\phi = {n_\phi \over 3} - 1. \label{wphiquin}
\end{equation}
Hence
\begin{equation}
\rho_\phi \propto a^{-3(1+w_\phi)}.
\end{equation}
If one can neglect the background energy $\rho_B$, then (\ref{H2}) 
yields
a simple differential equation for $a(t)$ which is solved as~:
\begin{equation}
a \propto t^{2/[3(1+w_\phi)]}.
\end{equation}
Since ${\dot \phi}^2 = (1+w_\phi) \rho_\phi$, one deduces that $\phi$
varies logarithmically with time. One then easily obtains from
(\ref{eomphi},\ref{H2}) that 
\begin{equation}
\phi = \phi_0 + {2 \over \lambda} \ln (t/t_0). \label{phi}
\end{equation}
and\footnote{ under the condition $\lambda^2 \le 6$ ($w_\phi \le 1$
since $V(\phi) \ge 0$).}
\begin{equation}
w_\phi = {\lambda^2 \over 3} - 1, \label{wphi}
\end{equation}
It is clear from (\ref{wphi}) that, for $\lambda$ sufficiently small,
the field $\phi$ can play the role of quintessence.

But the successes of the standard big-bang scenario indicate that
clearly $\rho_\phi$ cannot  have always dominated: it must have emerged
from the background energy density $\rho_B$. Let us thus now consider the
case where $\rho_B$ dominates.
It turns out that the solution just discussed with $\rho_\phi \gg
\rho_B$ and (\ref{wphi}) is a late time attractor \cite{Halliwell} 
 only if $\lambda^2 < 3(1+ w_B)$. If $\lambda^2 > 3(1+ w_B)$, the
global attractor turns out to be  a scaling solution \cite{Wett,CLW,FJ}
with the following properties:\footnote{See ref.\cite{JPU} for the case
where the scalar field is non-minimally coupled to gravity.}
\begin{eqnarray}
\Omega_\phi \equiv {\rho_\phi \over \rho_\phi+ \rho_B} &=& 
{3 \over \lambda^2} (1 + w_B) \label{Omegaphi} \\
w_\phi &=& w_B \label{wB}
\end{eqnarray}
The second equation (\ref{wB}) clearly indicates that this does not
correspond to a quintessence solution (\ref{eqw}). 

The semi-realistic models discussed earlier tend to give large values of 
$\lambda$ and thus the latter scaling solution as an attractor. For
example, in the case (\ref{larges}) where the scalar field is the
dilaton, 
$\lambda = 3/b$ with $b = C /(16 \pi^2)$ and $C= 90$ for a $E_8$ gauge
symmetry down to $C=9$ for $SU(3)$. Moreover \cite{FJ}, on the
observational side, the condition that $\rho_\phi$ should be
subdominant during nucleosynthesis (in the radiation-dominated era)
imposes to take rather large values of $\lambda$. Typically requiring
$\rho_\phi / (\rho_\phi + \rho_B)$ to be then smaller than $0.2$
imposes $\lambda^2 > 20$. 

Although not quintessence, such attractor
models with a fixed fraction $\Omega_\phi$ as in (\ref{Omegaphi}) have
interest of their own \cite{FJ}, 
in particular for structure formation if $\lambda
\in [5,6]$. It has been proposed recently \cite{DKS}  to make the
prefactor $V_0$ in (\ref{Vexp}) a trigonometric function in $\phi$. This
allows for some modulation around the previous attractor in an
approximately oscillatory way: $\Omega_\phi$ could then have been small
at the time of nucleosynthesis and be much larger at present times.
Finally, very recently \cite{HeWe}, such models have been coupled to a
system of a Brans-Dicke field and a dynamical field characterizing the
cosmological constant, with a diverging kinetic term, to provide a
relaxation mechanism for the cosmological constant \cite{Rubakov}.

Ways to obtain a quintessence component have been proposed however. 
Let me sketch 
some of them in turn.

One is the notion of {\em tracker field}\footnote{ Somewhat of a
misnomer since in this solution, as we see below, the field $\phi$
energy density tracks the radiation-matter energy
density before overcoming it  (in contradistinction with
(\ref{Omegaphi})). One should rather describe it as a {\em transient tracker field}.} \cite{ZWS}. This idea
also rests on the existence of scaling solutions of the equations of
motion which play the role of late time attractors, as illustrated
above. An example is provided by a scalar field described by
the action (\ref{action}) with a potential \begin{equation}
V(\phi) = \lambda \ {\Lambda^{4+\alpha} \over \phi^\alpha}
\label{trackerpot}
\end{equation}
with $\alpha >0$. In the case where the background density dominates,
one finds an attractor scaling solution \cite{RP,PR,LS,JPU} 
$\phi \propto
a^{3(1+w_B)/(2+\alpha)}$, $\rho_\phi \propto 
a^{-3\alpha (1+w_B)/(2+\alpha)}$. Thus $\rho_\phi$ decreases at a
slower rate than the background density ($\rho_B \propto a^{-3(1+
w_B)}$) and tracks it until it becomes of the same order at a given
value $a_Q$. More precisely \cite{B}~:
\begin{eqnarray}
\phi &=& m_P \sqrt{{\alpha (2+\alpha) \over 3(1+w_B)}}
\left({a \over a_Q} \right)^{3(1+w_B)/(2+\alpha)}, \label{phievol} \\
\rho_\phi &\sim & \lambda {\Lambda^{4+ \alpha} \over m_P^\alpha} 
\left( {a \over a_Q} \right)^{-3\alpha (1+w_B)/(2+\alpha)}.
\label{rhoevol}
\end{eqnarray}
One finds
\begin{equation}
w_\phi = -1 + {\alpha (1+ w_B) \over 2+ \alpha}. \label{wquin}
\end{equation}
Shortly after  $\phi$ has reached for $a=a_Q$ a value of order $m_P$, it 
satisfies the standard slowroll conditions:
\begin{eqnarray}
m_P |V' / V| &\ll & 1, \\
m_P^2 |V'' / V| &\ll & 1,
\end{eqnarray}
and therefore (\ref{wquin}) provides a good approximation to the
present value of $w_\phi$.
Thus, at the end of the matter-dominated era, this field may provide
the quintessence component that we are looking for. 

Two features are particularly interesting in this respect. One is that
this scaling solution is reached for rather general initial conditions,
{\em i.e.} whether $\rho_\phi$ starts of the same order or much smaller
than  the background energy density \cite{ZWS}. 
The second deals with the central question in this game: why is the
$\phi$ energy density (or in the case of a cosmological constant, the
vacuum energy density) emerging now? Since $\phi$ is of order $m_P$ in
this scenario, it can be rephrased here into the following: why is
$V(m_P)$ of the order of the critical energy density $\rho_c$? Using
(\ref{rhoevol}), this amounts to a constraint on the parameters of the
theory: \begin{equation}
\Lambda \sim \left( H^2_0 m_P^{2+\alpha} \right)^{1/(4+\alpha)} .  
\label{Lambdapar}
\end{equation}
For example, this gives for $\alpha = 2$, $\Lambda \sim 10$ MeV,
not such an unnatural value.

Let us note here the key difference between this tracking scenario and
the preceding one\footnote{ I wish to thank M. Joyce for discussions on
this point.}. Whereas the exponential potential model accounts for a
fixed fraction $\Omega_\phi$ in the attractor solution (and thus $\phi$
is a tracker in the strict sense), the final  attractor in
the tracker field solution corresponds to scalar field dominance
($\Omega_\phi \sim 1$). It is the scale $\Lambda$ which allows to
regulate the time at which the scalar field starts to emerge and makes it
coincide with present time. The welcome property is that the required
value for $\Lambda$ falls in a reasonable range from a high energy
physics point of view. 
On the other hand, we will see below that the fact that the
present value for $\phi$ is of order $m_P$ is a source of problems.

Models of dynamical supersymmetry breaking easily provide 
a model of the type just discussed \cite{B}. Let us consider
supersymmetric QCD with gauge group $SU(N_c)$ and $N_f<N_c$ flavors,
{\em i.e.} $N_f$ quarks $Q^i$ (resp. antiquarks $\bar Q_i$), $i=1 \cdots
N_f$,  in the fundamental ${\bf N_c}$ (resp. anti-fundamental ${\bf \bar
N_c}$ of $SU(N_c)$. At the scale of dynamical symmetry breaking
$\Lambda$ where the gauge coupling becomes strong\footnote{It is given by
an expression such as (\ref{Lambda}) where $g_0$ is the value of the
gauge coupling at the large scale $M_s$ and $b$ the one-loop beta
function coefficient for gauge group $SU(N_c)$.}, boundstates of the
meson type form~: $ {\Pi^i}_j = Q^i \bar Q_j$. The dynamics is described
by a superpotential which can be computed non-perturbatively using
standard methods~: \begin{equation}
W= (N_c-N_f) { \Lambda^{(3N_c-N_f)/(N_c-N_f)} \over \left( {\rm det} \
\Pi \right)^{1/(N_c-N_f)}}.
\end{equation}
Such a superpotential has been used in the past but with the addition of 
a mass or interaction term ({\em i.e.} a positive power of $\Pi$) in
order to stabilize the condensate. One does not wish to do that here if
$\Pi$ is to be interpreted as a runaway quintessence component. For
illustration purpose, let us consider a condensate diagonal in flavor
space: ${\Pi^i}_j \equiv \phi^2 \delta^i_j$ (see \cite{MPR} for a more
complete analysis). Then the potential for $\phi$ has the form
(\ref{trackerpot}), with $\alpha= 2 (N_c+N_f)/(N_c-N_f)$. Thus,
\begin{equation}
w_\phi = -1 + {N_c+N_f \over 2N_c} (1+ w_B),
\end{equation}
which clearly indicates that the meson condensate is a potential
candidate for a quintessence component.

Another possibility for the emergence of the quintessence component out
of the background energy density might be attributed to the presence of
a local minimum (a ``bump'') in the potential $V(\phi)$: when the field $\phi$
approaches it, it slows down and $\rho_\phi$ decreases more slowly
($n_\phi$ being much smaller as $w_\phi$ temporarily becomes closer to
$-1$, cf.(\ref{wphiquin})). If the parameters of the potential are chosen
carefully enough, this allows the background energy density, which
scales as $a^{-3(1+w_B)}$ to become subdominant. The value of $\phi$ at
the local minimum provides the scale which allows to regulate the time
at which this happens. This approach can be traced back to the earlier
work of Wtterich \cite{Wett} and has
recently been advocated by Albrecht and Skordis \cite{AS} in the
context of an exponential potential. They argue quite sensibly that, in
a ``realistic'' string model, $V_0$ in (\ref{Vexp}) is
$\phi$-dependent: $V_0(\phi)$. This new field dependence might be such
as to generate a bump in the scalar potential and thus a local minimum.
Since \begin{equation}
{1 \over V} {dV \over d\phi} = {V'_0 (\phi) \over V_0(\phi)} - {\lambda
\over m_P},
\end{equation}
it suffices that $m_P V'_0 (\phi)/ V_0(\phi)$ becomes temporarily
larger than $\lambda$ in order to slowdown the redshift of
$\rho_\phi$: once $\rho_\phi$ dominates, an attractor scaling solution
of the type (\ref{phi},\ref{wphi}) is within reach, if $\lambda$ is
not too large. As pointed out by Albrecht and Skordis, the success of
this scheme does not require very small couplings.

One may note that, in the preceding model, one could arrange the local
minimum in such a way as to completely stop the scalar field, allowing for a
period of true inflation \cite{PeRo}. The last possibility that I will
discuss goes in this direction one step further. It is known under
several names: deflation \cite{Spokoiny}, kination \cite{Joyce,JoPr},
quintessential inflation \cite{PV}. It is based on the remark that, if a
field $\phi$ is to provide a dynamical cosmological constant under the
form of quintessence, it is a good candidate to account for an
inflationary era where the evolution is dominated by the vacuum energy.
In other words, are the quintessence component and the inflaton  the
same unique field?

In this kind of scenario, inflation (where the energy density of
the Universe is dominated by the $\phi$ field potential energy) is
followed by reheating where matter-radiation is created by gravitational
coupling during an era where the
evolution is driven by the $\phi$ field kinetic energy (which decreases
as $a^{-6}$). Since matter-radiation energy density is decreasing more slowly,
this turns into a radiation-dominated era until the
$\phi$ energy density eventually emerges as in the quintessence scenarios
described above.

Finally, it is worth mentioning that, even though the models discussed above
all have $w_\phi \ge -1$, models with lower values of $w_\phi$ may easily be 
constructed. One may cite models with non-normalized scalar  field kinetic
terms \cite{COY}, or simply models with non-minimally coupled scalar fields
\cite{RiUz}. Indeed, it has been argued by Caldwell \cite{caldwell} that such 
a ``phantom'' energy density component fits well the present observational 
data. 

\subsection{Pseudo-Goldstone boson}

There exists a class of models \cite{FHSW} very close in spirit to the case of
runaway quintessence: they correspond to a situation where a scalar
field has not yet reached its stable groundstate and is still evolving in its potential. 

More specifically, let us consider a potential of the form:
\begin{equation}
V(\phi) = M^4 v \left( {\phi \over f} \right),
\end{equation}
where $M$ is the overall scale, $f$ is the vacuum expectation value
$<\phi>$ and the function $v$ is expected to have coefficients of
order one. 
If we want the potential energy of the field (assumed to be close to its
{\em vev} $f$) to give a substantial fraction of the energy density at
present time, we must set 
\begin{equation}
M^4 \sim \rho_c \sim H_0^2 m_P^2 . \label{PGBen}
\end{equation}
However, requiring that the evolution of the field $\phi$ around its minimum
has been overdamped by the expansion of the Universe until recently imposes
\begin{equation}
m_\phi^2 = {1 \over 2} V'' (f) \sim {M^4 \over f^2} \le H_0^2.
\label{PGBmass} \end{equation}
Let us note that this is one of the slowroll conditions familiar to the
inflation scenarios. 

From (\ref{PGBen}) and (\ref{PGBmass}), we conclude
that $f$ is of order $m_P$ (as the value of the field $\phi$ in runaway
quintessence) and that $M \sim 10^{-3}$ eV (not surprisingly, this is
the scale $\Lambda$ typical  of the cosmological constant, see 
(\ref{Lambdaconstr})). The field $\phi$ must be very light: $m_\phi \sim
h_0 \times 10^{-60} m_P \sim h_0 \times 10^{-33} \; {\rm eV}$. Such a
small value is only natural in the context of an approximate symmetry:
the field $\phi$ is then a pseudo-Goldstone boson.

A typical example of such a field is provided by the axion field
(QCD axion \cite{Kim} or string axion \cite{Choi}). In this case, the
potential simply reads~:
\begin{equation}
V(\phi) = M^4 \left[ 1+ \cos (\phi / f) \right].
\end{equation}

 \section{Quintessential problems}

However appealing, the quintessence idea is difficult to implement in
the context of realistic models  \cite{Carroll,KL}. The main problem lies in
the fact that the quintessence field must be extremely weakly coupled
to ordinary matter. This problem can take several forms~:

\vskip .3cm $\bullet$
we have assumed until now that the quintessence potential
monotonically decreases to zero at infinity. In realistic cases, this
is difficult to achieve because the couplings of the field to ordinary
matter generate higher order corrections that are increasing with larger
field values, unless forbidden by a symmetry argument.  For example, in
the case of the potential (\ref{trackerpot}), the generation of  a
correction term $\lambda_d \ m_P^{4-d}\phi^d$ puts in jeopardy the slowroll
constraints  on the quintessence field,
unless very stringent constraints are imposed on the coupling
$\lambda_d$. But one typically expects from supersymmetry breaking 
$\lambda_d \sim M_S^4/m_P^4$ where $M_S$ is the supersymmetry breaking scale
\cite{KL}. 

Similarly, because the $vev$ of $\phi$ is of order $m_P$, one must take into 
account the full supergravity corrections. One may then argue \cite{BrMa} 
that this could put in jeopardy the positive definiteness of the scalar 
potential, a key property of the quintessence potential. This may point 
towards models where $<W> = 0$ (but not its derivatives, see (\ref{sugrapot}))
or to no-scale type models: in the latter case, the presence of 3 moduli 
fields $T^i$ with K\"ahler potential $K= - \sum_i \ln (T^i + \bar T^i)$ 
cancels the negative contribution $-3 |W|^2$ in (\ref{sugrapot}).\footnote{
Moreover, supergravity corrections may modify some of the results. For example,
the presence of a (flat) K\"ahler potential in (\ref{sugrapot}) induces 
exponential field-dependent factors. A more adequate form for the inverse 
power law potential (\ref{trackerpot}) is thus \cite{BrMa} $V(\phi) 
= \lambda e^{\phi^2  / 2 M_P^2} \Lambda^{4+ \alpha} / \phi^\alpha$. 
The exponential factor is not 
expected to change much the late time evolution of the quintessence energy 
density. Brax and  Martin \cite{BrMa} argue that it changes the equation of 
state through the  value of $w_\phi$.}

\vskip .3cm $\bullet$ 
the quintessence field must be very light \cite{Carroll}. If we
return to our example of supersymmetric QCD in (\ref{trackerpot}), 
$V''(m_P)$ provides an order of magnitude for the mass-squared  of the 
quintessence component:
\begin{equation}
m_\phi \sim \Lambda \left( { \Lambda \over m_P } \right)^{1+\alpha/2}
\sim H_0 \sim 10^{-33} \ {\rm eV}.
\end{equation}
using (\ref{Lambdapar}). 
Similarly, we have seen that the mass of a pseudo-Goldstone boson that could 
play the r\^ole of quintessence is typically of the same order. This 
field must therefore be very weakly coupled to matter; otherwise its
exchange would generate observable long range forces. E\"otv\"os-type 
experiments put very severe constraints on such couplings.

Again, for the case of supersymmetric QCD, higher order corrections to the 
K\"ahler potential of the type
\begin{equation}
\kappa(\phi_i, \phi_j^\dagger) \left[ \beta_{ij} \left( {Q^\dagger Q \over 
m_P^2} \right) 
+ \bar \beta_{ij} \left( {\bar Q \bar Q^\dagger  \over 
m_P^2} \right) \right] \label{ho}
\end{equation}
will generate couplings of order 1 to the  standard matter fields $\phi_i$,
$\phi^\dagger_j$ since $<Q>$ is of order $m_P$. In order to alleviate this problem, 
Masiero, Pietroni and Rosati \cite{MPR} have proposed a solution much in the 
spirit of the least coupling principle of Damour and Polyakov \cite{DaPo}: 
the different functions $\beta_{ij}$ have a {\em common} minimum close to the 
value $<Q>$, which is most easily obtained by assuming ``flavor'' independence 
of the functions $\beta_{ij}$. This obviously eases the E\"otv\"os experiment 
constraints. In the early stages of the evolution of the Universe, when 
$Q \ll m_P$, couplings of the type (\ref{ho}) generate a contribution to the 
mass of the $Q$ field which, being proportional to $H$, does not spoil the 
tracker solution.

\vskip .3cm $\bullet$ 
it is difficult to find a symmetry that would prevent any
coupling of the form $\beta (\phi/m_P)^n F^{\mu \nu} F_{\mu\nu}$ to the
gauge field kinetic term. Since the quintessence behavior is
associated with time-dependent values of the field of order $m_P$, this
would generate, in the absence of fine tuning, corrections of order one
to the gauge coupling. But the time dependence of the fine structure
constant for example is very strongly constrained \cite{DaDy}: $|\dot \alpha /
\alpha| < 5 \times 10^{-17} {\rm yr}^{-1}$. This yields a limit \cite{Carroll}:
\begin{equation}
|\beta | \le 10^{-6} {m_P H_0 \over <\dot \phi>},
\end{equation}
where $<\dot \phi >$ is the average over the last $2 \times 10^9$ years.

A possible solution is to implement an approximate continuous symmetry of the 
type: $\phi \rightarrow \phi + $ constant \cite{Carroll}. This symmetry must 
be approximate since it must allow for a potential $V(\phi)$. Such a symmetry 
would only allow derivative couplings, an example of which is an axion-type 
coupling $\tilde \beta (\phi / m_P ) F^{\mu\nu} \tilde F_{\mu \nu}$. If 
$F_{\mu\nu}$ is the color $SU(3)$ field strength, QCD instantons yield a 
mass of order $\tilde \beta \Lambda^2_{_{QCD}}/m_P$, much too large to satisfy 
the preceding constraint. In any case, supersymmetry would relate such a 
coupling to the coupling  $\beta (\phi / m_P ) F^{\mu\nu} F_{\mu \nu}$ that 
we started out to forbid.

\vskip .3cm

The very light mass of the quintessence component points towards scalar-tensor 
theories of gravity, where such a dilaton-type (Brans-Dicke) scalar field is 
found. This has 
triggered some recent interest for this type of theories. Attractor scaling  
solutions 
have been found for non-minimally coupled fields \cite{JPU,Amen}. However, as 
discussed above, one problem is that scalar-tensor theories lead to 
time-varying constants of nature. One may either put some limit on the 
couplings of the scalar field \cite{Chiba} or use the attractor mechanism 
towards General Relativity that was found by Damour and Nordtvedt 
\cite{DaNo,DaPo}. This mechanism exploits the stabilsation of the dilaton-type
scalar through its conformal coupling to matter. Indeed, assuming that this
scalar field $\phi$ 
couples to matter through an action term ${\cal S}_m \left(\psi_m, f(\phi) 
g_{\mu\nu}) \right)$, then its equation of motion takes the form:
\begin{equation}
{2 \over 3-{\phi'}^2} \phi'' + (1-w_B) \phi' = - (1-3w_B) {d \ln f(\phi) 
\over d\phi},
\end{equation}
where $\phi' = d \phi / d \ln a$.
This equation can be interpreted \cite{DaNo} as the motion of a particle of 
velocity-dependent mass $2/(3-{\phi'}^2)$ subject to a damping force
$(1-w_B) \phi'$ in an external force deriving from a potential
$v_{{\rm eff}} (\phi) = (1-3w_B) \ln f(\phi)$. If this effective potential 
has a minimum, the field quickly settles there.
Bartolo and Pietroni \cite{BaPi} have recently proposed a model of quintessence
(they add a potential $V(\phi)$) using this mechanism: 
the quintessence component is first attracted to General Relativity and then 
to a standard tracker solution.

Scalar-tensor theories of gravity naturally arise in the context of 
higher-dimensional theories and we will return to such scenarios in the next 
section where we discuss these theories.

All the preceding shows that there is extreme fine tuning in the
couplings of the quintessence field to matter, unless they are
forbidden by some symmetry. This is somewhat reminiscent of the fine
tuning associated with the cosmological constant. In fact, {\em  the
quintessence solution does not claim to solve the cosmological
constant (vacuum energy) problem} described above. If we take the
example of a supersymmetric theory, the dynamical cosmological
constant provided by the quintessence component clearly  does not provide
enough amount of supersymmetry breaking to account for the mass
difference between scalars (sfermions) and fermions (quarks and
leptons): at least $100$ GeV. There must be other sources of
supersymmetry breaking and one must fine tune the parameters of the
theory in order not to generate a vacuum energy that would completely
drown $\rho_\phi$.

In any case, the quintessence solution shows that, once this fundamental
problem is solved, one can find explicit fundamental models that
effectively provide the small amount of  cosmological constant that
seems required by experimental data.

\section{Extra spacetime dimensions}

The old idea of Kaluza and Klein about compact extra dimensions has received a 
new twist with the realisation, motivated by string theory \cite{HoWi}, that 
such extra dimensions  may only be felt by gravitational interactions 
\cite{ADD}. In other words, our 4-dimensional world of quarks, leptons 
and gauge interactions may constitute a hypersurface in a 
higher-dimensional Universe.
Such a hypersurface is called a brane in modern jargon:  certain types of
branes (Dirichlet branes) appear as solitons in open string theories 
\cite{Pol}. In what follows, we will mainly consider 4-dimensional branes 
to which  are confined observable matter as well as standard non-gravitational 
gauge interactions. The part of the Universe which is not confined to the brane
is called the bulk (which for simplicity we will take to be 5-dimensional).

In this framework, the very notion of a cosmological constant takes a new  
meaning and there has been recently a lot of activity to try to unravel it. The
hope is that the cosmological constant problem itself may receive a different 
formulation, easier to deal with.

If we think of the cosmological constant as some vacuum energy, one has the
choice to add it 
to the brane or to the bulk. The consequences are quite different:

\vskip .3cm
$\bullet$
If we introduce a vacuum energy $\lambda_b>0$ on the brane, it creates 
a repulsive gravitational force outside ({\em i.e.} in the bulk). Indeed, 
a result originally obtained by Ipser and Sikivie \cite{IpSi} in the 
case of a domain wall may be adapted here as follows: let $p$ and $\rho$ be the
pressure and energy density on the brane, then if $\rho + 3 p$ is positive
({\em resp.} negative), a test body may remain in the bulk stationary to the 
brane if it accelerates away from ({\em resp.} towards) the brane.\footnote{
One may note that, if the expansion in the brane is standard, then, according 
to (\ref{accdec}), the expansion in the brane is decelerating ({\em resp.}
accelerating).}  In the case 
of a positive cosmological constant, $\rho =-p = \lambda_b$ and 
$\rho + 3 p = -2 \lambda_b < 0$.

Projected back to our 4-dimensional brane-world, this yields a different 
behaviour from the one seen in a standard 4-dimensional world. For example, the
vacuum energy contributes to the Hubble parameter describing the
expansion of the brane world in a (non-standard) quadratic way \cite{BDL}: 
$H^2 = \lambda^2_b / (36 M^6) + \cdots$, where $M$ is the fundamental 
5-dimensional scale.  

\vskip .3cm
$\bullet$
If we introduce a vacuum energy $\lambda_B$ ) in the 5-dimensional bulk 
(this $\lambda_B$ is then of mass dimension 5), 
this will induce a potential for the modulus field whose {\em vev} 
measures the radius of the compact dimension. 
Let us call  for simplicity $R$ this modulus, which 
is often referred to as the radion. Then in the case of a single 
compact dimension, $V(R) = \lambda_B R$ \cite{AHDMR}.

The contribution of this bulk vacuum energy to the square of the Hubble 
parameter on the 
brane is standard (linear)~: $H^2 = \lambda_B / (6 M^3) + \cdots$

\vskip .3cm
Allowing both types of vacuum energies allows to construct static solutions 
with a cancelling effect in the bulk. Indeed, if one imposes the condition~:
\begin{equation}
\lambda_B = - {\lambda_b^2 \over 6 M^3},  \label{RS} 
\end{equation}
the effective 4-dimensional cosmological constant, {\em i.e.} the constant 
term in the Hubble parameter $H$, vanishes.

A striking property of this type of configuration is that it allows to 
localize gravity on the brane. This is the so-called Randall-Sundrum scenario
\cite{RaSu} (see also  \cite{prec} for earlier works). The 5-dimensional 
Einstein equations are found to allow for a 4-dimensionally flat solution 
with a  warp factor ({\em i.e.} an overall fifth dimension-dependent factor 
in front  of the four-dimensional metric)~:
\begin{equation}
ds^2 = e^{-|y\lambda_b|/(3M^3)} \eta_{\mu\nu} dx^\mu dx^\nu + dy^2
\end{equation}
if the condition (\ref{RS}) is satisfied. Let us note that this condition 
ensures that the bulk is anti-de Sitter since $\lambda_B <0$. If $\lambda_b>0$,
one finds a  single {\em normalisable} massless mode of the metric which is 
interpreted as the massless 4-dimensional graviton. The wave function of this 
mode turns out to be localized close to the brane, which gives an explicit 
realisation of 4-dimensional gravity trapping.   There is also a continuum
of non-normalisable massive modes (starting from zero mass) which 
are interpreted as the Kaluza-Klein graviton modes.

Of course, the Randall-Sundrum condition (\ref{RS}) is another version of the 
standard fine tuning associated with the cosmological constant. One would like 
to find a dynamical justification to it. 

Some progress has recently been made in this direction \cite{ADKS,KaScSi}.
The presence of a scalar field in the bulk, conformally coupled to the matter 
on the brane allows for some relaxation mechanism that screens the 
4-dimensional cosmological constant from corrections to the brane vacuum 
energy. Let us indeed consider such a scalar field, of the type discussed 
above in the context of scalar-tensor theories. The action is of the following 
form~:
\begin{eqnarray}
{\cal S} &=& \int d^5x \sqrt{g^{(5)}} \left[ {M^3 \over 2} R^{(5)} 
- {1 \over 2} \partial^N \phi \ \partial_N \phi - V(\phi) \right] \nonumber \\
& &+ \; {\cal S}_m \left(\psi_m, g_{\mu\nu} f(\phi)\right)
\end{eqnarray}
where  the fields $\psi_m$ are matter fields localized on the brane, located
at $y=0$, and 
$g_{\mu\nu}$ is the 4-dimensional metric ($N$ are 5-dimensional indices 
whereas $\mu$, $\nu$ are 4-dimensional indices). We will be mostly interested 
in the 4-dimensional vaccuum energy so that we can write the 4-dimensional 
matter action as~:
\begin{equation}
{\cal S}_m = -\int d^4 x \sqrt{g^{(4)}} \lambda_b f^2(\phi).
\end{equation}

 Five-dimensional  Einstein equations 
projected on the brane, provide the following Friedmann equation:
\begin{equation}
H^2 =  {1 \over 18 M^6} \lambda_b^2 f^2(\phi) 
\left[ f^2(\phi) - 3M^2 {f'}^2(\phi) \right] 
+ {1 \over 3} V(\phi).
\end{equation}
The other equations, including the $\phi$ equation of motion, ensure that this 
vanishes, irrespective of the precise value of $\lambda_b$, for the following metric~:
\begin{equation}
ds^2 = e^{-\alpha(y)} dx^\mu dx^\nu + dy^2,
\end{equation}
where the derivative of the function $\alpha(y)$ with respect to $y$ is fixed 
on the brane by junction conditions (assuming a symmetry $y \rightarrow -y$)
\begin{equation}
\alpha'(0) = {\lambda \over 3 M^5} f^2(\phi)|_{{\rm y=0}}.
\end{equation} 
In other words, 
the cosmological constant is, to a first order, not sensitive to the 
corrections to the vacuum energy coming from the Standard Model interactions. 

For specific values of the potential, such a dynamics localizes the gravity
around the brane. For example, with vanishing potential, the solution of the 
equations is obtained for 
\begin{equation}
f(\phi) = e^{\phi/(M\sqrt{3})}.
\end{equation}
One obtains a flat 4-dimensional spacetime (indeed, in this case, this
is the unique solution \cite{ADKS}) although the vacuum energy may 
receive non-vanishing corrections. 

The price to pay is the presence of a singularity close to the brane. It 
remains to be seen what is the interpretation of this singularity, how it 
should be treated and whether this reintroduces fine tuning 
\cite{DM2000,FLLN,CEGH}. Also a full cosmological treatment, {\em i.e.} including 
time dependence, is needed.

Presumably, supersymmetry plays an important role in this game if one wants
to deal with stable solutions. Supersymmetry indeed may prove to be in the 
end the rationale for the vanishing of the cosmological constant. The picture 
that would emerge would be one of a supersymmetric bulk with vanishing 
cosmological constant and with supersymmetry broken on the brane (remember that
supersymmetry is related to translational invariance) \cite{verlinde}.
Models along these lines have been discussed recently by Gregory,
Rubakov and Sibiryakov \cite{GRS}: the four-dimensional gravity is
localized on the brane due to the existence of an unstable graviton
boundstate.
Presumably in such models one does not recover the standard theory of gravity.

\section{Conclusion}

The models discussed above are many. This is not a surprise since the 
cosmological constant problem, although it has attracted theorists for decades,
has not received yet a convincing treatment. What is 
new is that one expects in a not too distant  future a large and 
diversified amount 
of observational data that should allow to discreminate among these models.
One may mention the MAP and PLANCK satellites on the side of CMB measurements.
The SNAP mission should provide, on the other hand, large numbers of type Ia
supernovae which should allow a better handle on this type of measurements 
and a significant increase in precision. But other methods will also give 
complementary information: lensing, galaxy counts \cite{NeDa}, gravitational 
wave detection \cite{Giovannini,RiUz}, ...

\vskip 2cm
{\bf Acknowledgments:}
\vskip 1cm
I wish to thank Christophe Grojean, Michael Joyce, Reynald Pain, James
Rich and Jean-Philippe Uzan for discussions and 
valuable  comments on the manuscript. I thank the Theory Group of Lawrence 
Berkeley National Lab where I found ideal conditions to finish writing these 
lecture notes.

\end{document}